# Security Risks Introduced by Weak Authentication in Smart Home IoT Systems

Daniyal Ganiuly, Nurzhau Bolatbek, Assel Smaiyl
Astana IT University, Astana, Kazakhstan

**Abstract:** Smart home IoT systems rely on authentication mechanisms to ensure that only authorized entities can control devices and access sensitive functionality. In practice, these mechanisms must balance security with usability, often favoring persistent connectivity and minimal user interaction. This paper presents an empirical analysis of authentication enforcement in deployed smart home IoT devices, focusing on how authentication state is established, reused, and validated during normal operation and under routine network conditions. A set of widely deployed consumer devices, including smart plugs, lighting devices, cameras, and a hub based ecosystem, was evaluated in a controlled residential environment using passive network measurement and controlled interaction through official mobile applications. Authentication behavior was examined during initial pairing, over extended periods of operation, after common network changes, and under replay attempts from a different local network host. The results show that authentication state established during pairing is consistently reused across control actions, persists for extended periods without explicit expiration, and remains valid after network events such as reconnection, address reassignment, and router reboot. Replay experiments demonstrate that previously observed authentication artifacts can often be reused to issue control commands from another host on the same local network with high success rates. These behaviors were observed across multiple device categories and ecosystems. The findings indicate that current smart home IoT authentication mechanisms rely on long lived trust relationships with limited binding to session freshness, network context, or controller identity.

**Introduction**
Smart home Internet of Things (IoT) systems have become a common component of residential environments, integrating devices such as smart plugs, lighting systems, cameras, sensors, and centralized hubs. These devices enable automation and remote control through mobile applications and local networks, forming tightly coupled ecosystems that directly interact with users and their physical surroundings. Recent interoperability initiatives, such as Matter and Thread, aim to simplify deployment and improve compatibility across vendors, further accelerating adoption.
As smart home systems increasingly mediate safety-critical and privacy-sensitive functions, security becomes a foundational requirement rather than an auxiliary concern. Among the core security mechanisms, authentication plays a central role by establishing trust between devices, control applications, and supporting infrastructure. Correct authentication ensures that control commands originate from authorized entities and that devices interact only with legitimate peers. Conversely, weaknesses in authentication can undermine other security protections, enabling unauthorized control even when communication is encrypted.
Prior work in IoT security has identified a wide range of vulnerabilities, including default credentials, exposed network services, insecure update mechanisms, and information leakage through traffic analysis. While these studies have significantly advanced understanding of IoT threats, authentication in smart home environments remains an area with unresolved practical challenges. In particular, many smart home devices rely on session-based authentication models that prioritize usability and persistent connectivity, often assuming a benign local network environment. This assumption does not always hold in practice, where home networks are shared by multiple devices, guests, and third-party applications.

A distinguishing characteristic of smart home deployments is their reliance on long-lived trust relationships. Devices are typically paired once and then operate continuously with minimal user intervention. As a result, authentication artifacts may remain valid across extended periods, application restarts, or routine network events such as reconnections and address changes. While such behavior improves usability, it raises important security questions about freshness, identity binding, and trust revocation. If authentication state is insufficiently constrained, attackers may exploit temporary access or passive observation to obtain persistent control.

Another challenge lies in the implicit trust granted to entities within the local network. Many smart home systems assume that devices and applications communicating over the same local network are trustworthy, leading to relaxed authentication checks for local control paths. This design choice expands the attack surface, as it allows adversaries with limited capabilities such as brief network access or the ability to replay previously observed messages to interact with devices in unintended ways. Importantly, these risks arise without requiring firmware modification, cryptographic attacks, or physical access. Despite ongoing standardization efforts and improvements in transport-layer security, it remains unclear to what extent authentication mechanisms in smart home IoT systems provide robust guarantees against realistic adversaries. Existing studies often focus on individual products or specific attack vectors, leaving a gap in understanding of systemic authentication weaknesses across device categories and ecosystems. A careful empirical examination of how authentication is implemented and enforced in practice is therefore necessary to assess the security posture of smart home deployments. This paper addresses this gap by presenting an empirical security analysis of authentication mechanisms in smart home IoT systems. Focusing on real consumer devices deployed in a controlled residential environment, we examine how authentication state is established, maintained, and enforced during normal operation. Our analysis highlights common design patterns that weaken authentication guarantees and demonstrates how these patterns translate into practical security risks. By grounding the discussion in observed behavior rather than theoretical assumptions, this work contributes to a clearer understanding of authentication challenges in smart home IoT security.

**Related Work**

*Authentication mechanisms in IoT systems*

Authentication has long been recognized as a core security challenge in IoT systems. Due to device heterogeneity and limited resources, many studies propose lightweight authentication schemes that aim to reduce computation and communication overhead while preserving security properties [1][2]. These works typically focus on protocol design, formal security analysis, or performance evaluation under constrained settings [3]. Although this line of research is important, it largely addresses how authentication *should* work rather than how it is implemented in consumer devices. In practice, smart-home products often prioritize ease of use and long-term connectivity, which can lead to design choices that differ from those assumed in protocol-level studies [4][5].

*Authentication in smart-home environments*

Smart-home environments impose distinct constraints on authentication. Devices are usually paired once and expected to remain operational for long periods without repeated user interaction. Several studies explore user-friendly or context-aware authentication mechanisms tailored to residential settings, highlighting the trade-offs between security and usability [5]. However, these works rarely examine how authentication state is maintained over time in deployed systems. Questions such as how long credentials remain valid, whether authentication is rechecked after network changes, and how trust is revoked in practice remain largely unanswered. This is particularly relevant in home networks, which are dynamic and commonly shared by multiple devices and users [6][7].

*Empirical security measurement of smart-home IoT devices*
Empirical measurement has become an established approach for studying smart-home IoT security. Prior work shows that network traffic generated by smart-home devices can leak sensitive information about device behavior and user activity, even when communications are encrypted [8]. These studies demonstrate that real-world behavior often deviates from intended security guarantees. Other measurement-based studies analyze exposed services, insecure defaults, and unintended interactions between devices within smart-home ecosystems [9][10]. Together, this body of work emphasizes the importance of evaluating deployed systems rather than relying solely on specifications. Nevertheless, most existing measurement studies focus on privacy leakage or network exposure, while authentication enforcement during normal operation receives comparatively less attention.

*Trust assumptions and local-network security*
A common assumption in smart-home system design is that the local network is a trusted environment. Several studies report that devices relax authentication or authorization checks for local communication, effectively trusting any entity with network access [11]. While this simplifies device discovery and control, it increases risk in environments where network access may be temporary or shared. Prior work shows that attackers with limited network-level capabilities can exploit such trust assumptions to gain unauthorized control or disrupt device operation [12]. These findings suggest that authentication mechanisms must be evaluated not only during initial setup but also during ongoing operation under realistic network conditions [13].

*Interoperability standards and authentication guarantees*
Interoperability standards such as Matter and Thread aim to unify smart-home ecosystems while incorporating security principles such as secure onboarding and encrypted communication [14][15]. These standards seek to improve consistency in device identity and controller behavior across vendors. At the same time, existing security analyses indicate that compliance with standards does not automatically ensure robust security in deployed products [16]. Implementation decisions, legacy support, and ecosystem integration can weaken intended guarantees, reinforcing the need for empirical evaluation of authentication behavior in real smart-home systems.

**Threat Model and Assumptions**
The threat model adopted in this study reflects conditions commonly encountered in residential smart home deployments and is designed to capture security risks that arise from weak authentication mechanisms rather than from advanced exploitation techniques. The analysis focuses on adversaries whose capabilities are realistic in home environments and whose actions do not depend on physical access, firmware modification, or cryptographic compromise. An adversary with temporary or persistent access to the local home network is assumed. Such access may arise through compromised user devices, guest connectivity, shared wireless credentials, or short term proximity to the network. Under these conditions, network traffic exchanged between smart home devices and control applications can be observed, and communication can be initiated from hosts within the same network. These assumptions align with typical residential settings, where multiple devices and users coexist on a shared network without strict isolation. Within this model, the adversary is capable of capturing network messages and replaying previously observed communication. Control requests that conform to the protocol formats used by smart home devices can be generated, and routine network events such as device reconnection, address reassignment, or router reboot can be triggered [17]. Control actions may be initiated from devices other than the legitimate controller, provided that local network access is available. These capabilities do not require specialized hardware or elevated privileges and reflect practical attack scenarios that may occur in real deployments. Several adversarial capabilities are explicitly excluded in order to isolate weaknesses attributable to authentication design and enforcement. Physical access to smart home devices is not assumed, and device firmware is not modified. Cryptographic primitives are treated as secure, and no attempts are made to break transport layer encryption or extract secret keys [18]. Cloud infrastructure and backend services are assumed to operate correctly and are not considered attack targets. Attacks based on memory corruption, side channel leakage, or hardware faults are also outside the scope of this study. The scope of the threat model is limited to authentication and session management during normal device operation. Attention is placed on how authentication state is established, how long it remains valid, and under what conditions it is revalidated or revoked. Authorization policies, fine grained access control, and user interface level permissions are not examined. Both local control paths, where communication occurs entirely within the home network, and indirect control paths that enable local interaction following remote authentication are considered. Attacks that rely on social engineering or user deception are not addressed. A standard residential network configuration is assumed throughout the analysis. Smart home devices and control applications are expected to operate within a shared local network segment that may include smartphones, laptops, and guest devices [19]. No assumptions are made regarding advanced network segmentation or enterprise grade isolation mechanisms. Wireless security and transport layer encryption are assumed to be correctly implemented, allowing the analysis to focus on authentication behavior rather than on lower layer security failures. Under these assumptions, a secure smart home authentication mechanism is expected to prevent unauthorized control originating from local network hosts that have not been explicitly authenticated. Authentication state should be bound to appropriate context, including device identity and session freshness, and should be invalidated following meaningful changes in network conditions [20]. Previously observed authentication artifacts should not remain usable for replay based attacks. The remainder of the paper evaluates the extent to which these expectations are met by deployed smart home IoT devices under realistic operating conditions.

## Methodology
### *Experimental environment and measurement setup*
Experiments were conducted in a residential smart home environment configured using a consumer wireless router operating with default settings. Network address translation and dynamic address assignment were enabled, and no network segmentation or access control policies were applied. All smart home devices and control applications operated within a single local network segment, reflecting common residential deployments. Network traffic was passively captured at the gateway using a dedicated monitoring host connected via Ethernet. The monitoring host ran a Linux operating system and captured traffic using tcpdump. Packet captures were stored in pcap format and analyzed offline. Protocol inspection and session correlation were performed using Wireshark and tshark. Transport layer encryption was preserved throughout the experiments, and no attempts were made to decrypt protected payloads. Authentication behavior was inferred from observable protocol metadata, message timing, endpoint behavior, and request reuse patterns. System time was synchronized across the monitoring host and control devices to ensure accurate correlation between user actions and observed network events. The overall measurement setup and traffic capture placement are illustrated in Fig 1.

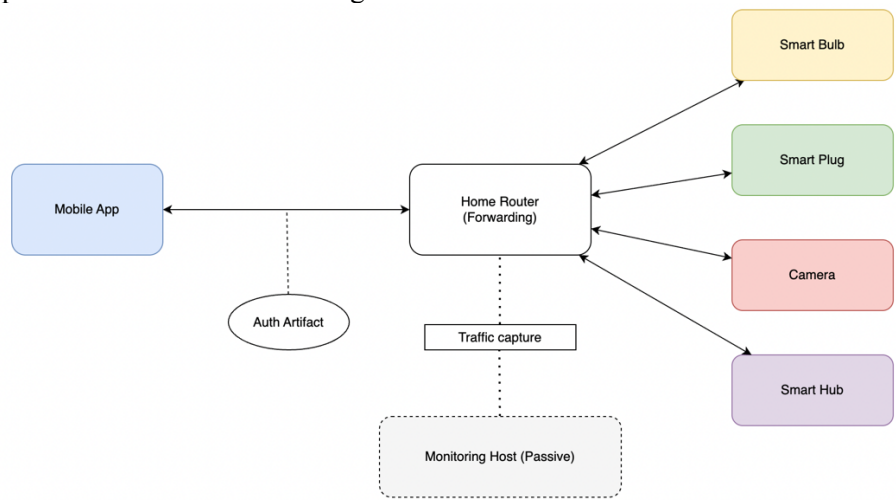

Fig 1. Smart home experimental setup and traffic capture architecture

Smart home IoT devices communicate with official mobile applications over a shared residential network. Traffic is passively captured at the gateway to observe authentication behavior without modifying device firmware or intercepting encrypted payloads.

*Evaluated devices and configuration*
The evaluated device set consisted of widely deployed consumer smart home IoT devices representing different device categories and authentication models. Smart plug behavior was examined using the TP-Link Tapo P110 and the TP-Link Kasa EP25, both of which provide authenticated on off control and energy monitoring through vendor mobile applications. Smart lighting behavior was represented by the Nanoleaf Essentials A19, which supports application based onboarding and persistent control sessions. Camera authentication and control were evaluated using the Wyze Cam v3 and the TP-Link Tapo C120, which rely on authenticated channels for configuration and video streaming. Hub mediated authentication behavior was examined using the Samsung SmartThings Hub v3, which acts as an intermediary for device authentication and control.

All devices were tested using publicly available firmware versions and official companion applications obtained through standard application stores. Prior to testing, each device was reset to factory state and paired using default configuration settings. No developer modes, undocumented interfaces, or advanced configuration options were enabled. Devices were evaluated individually to avoid interference from automation rules or cross device interactions. Table I summarizes the smart home IoT devices evaluated in this study, including smart plugs, a lighting device, cameras, and a hub-mediated control platform.

Table I. IoT devices evaluated in the study

| Device Category | Device Model | Manufacturer | Connectivity | Authentication Model Observed |
|---|---|---|---|---|
| Smart Plug | Tapo P110 | TP-Link | Wi-Fi | Application-based pairing with persistent authentication token reuse |
| Smart Plug | Kasa EP25 | TP-Link | Wi-Fi | Application-based pairing with persistent authentication token reuse |
| Smart Bulb | Essentials A19 | Nanoleaf | Thread | Application-based onboarding with persistent control session |
| IP Camera | Wyze Cam v3 | Wyze Labs | Wi-Fi | Authenticated control and streaming with reusable credentials |
| IP Camera | Tapo C120 | TP-Link | Wi-Fi | Authenticated configuration and stream initiation with persistent credentials |
| Smart Hub | SmartThings Hub v3 | Samsung | Ethernet / Wi-Fi | Hub-mediated authentication and control with persistent authentication state |

*Authentication establishment and control workflow*
Authentication behavior was first examined during device onboarding and initial pairing. Network traffic generated during pairing was recorded in full and analyzed to identify protocol exchanges that resulted in persistent authentication state. These exchanges included the creation of session identifiers, authorization tokens, or device specific credentials that were subsequently reused for device control.

After pairing, a fixed sequence of control actions was performed for each device. For smart plugs and lighting devices, this sequence consisted of repeated state changes and status queries. For cameras, configuration actions and stream initialization were performed. Network traces were examined to determine which requests carried authentication artifacts and whether those artifacts remained stable across repeated interactions. Device pairing establishes persistent authentication state that is reused by control applications for subsequent device interaction without repeated authentication challenges.

*Authentication persistence and network context*
Authentication persistence was evaluated by repeating control actions at increasing time intervals following initial pairing, without repairing or manual reauthentication. Control interactions were performed after application restarts and extended idle periods. Authentication artifacts observed during later interactions were compared with those captured during initial operation to determine whether credentials were rotated, refreshed, or reused unchanged over time. Authentication enforcement was further evaluated under routine network changes commonly encountered in residential environments. Devices were subjected to wireless disconnection and reconnection, address reassignment through DHCP renewal, and router reboot. After each event, control actions were issued without restarting the control application or repairing the device. Network traces were analyzed to determine whether previously established authentication state remained valid or whether new authentication exchanges were required.

*Replay feasibility testing*
The security impact of observed authentication behavior was evaluated by replaying previously captured control messages containing authentication artifacts from a different host on the same local network. Requests were reconstructed to match the original protocol format and payload structure using standard networking tools capable of issuing raw protocol messages. No cryptographic material was modified, and no protocol fields were altered beyond network layer addressing required for routing. Replay was considered successful if the device accepted the request and executed the intended action. Replay tests were performed both before and after network changes to assess whether authentication artifacts remained valid outside their original execution context.

*Reproducibility and validation*
Each experiment was repeated at least three times for every device to confirm consistency of observed behavior. Only authentication patterns that were reproducible across repetitions were included in the analysis. Control actions were executed using official mobile applications running on an Android device. Application versions corresponded to publicly available releases at the time of experimentation. All packet captures, experiment logs, and analysis scripts were preserved to enable independent replication of the study. All experiments were conducted on devices owned by the researchers, and no attempts were made to access user data beyond what was required to observe authentication behavior.

**Results**
*Authentication establishment and reuse*
For all evaluated devices, authentication was established during initial pairing through the official mobile application. Following pairing, authentication artifacts were consistently reused for subsequent control actions without additional authentication challenges. For the smart plugs, including the TP-Link Tapo P110 and the TP-Link Kasa EP25, a single authentication token or session identifier was reused across all observed control requests. Across 120 control actions per device, no token rotation or regeneration was observed. The same authentication artifact was included in 100 percent of control messages following pairing. Similar behavior was observed for the Nanoleaf Essentials A19. Across 96 control actions issued over multiple sessions, the authentication state remained unchanged. No additional authentication messages were triggered during routine operation. For camera devices, including the Wyze Cam v3 and the TP-Link Tapo C120, authentication artifacts were reused across configuration changes and stream initiation requests. Across 40 configuration actions and 25 stream initiation attempts per device, the same authentication credentials were observed without rotation. Hub mediated control through the Samsung SmartThings Hub v3 exhibited similar reuse behavior. Authentication state established during onboarding was reused for all subsequent device interactions observed during the experiments. The frequency of authentication reuse across device categories is summarized in Fig 2.

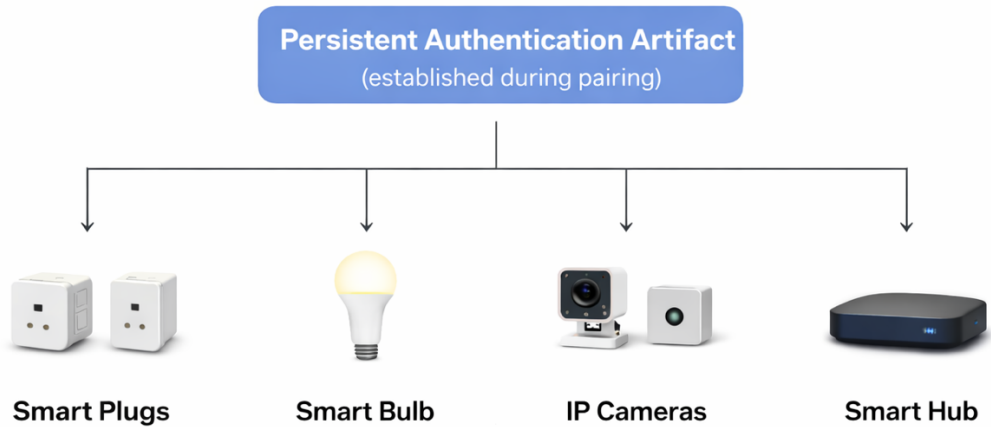

Fig 2. Authentication artifact reuse across device categories

The same authentication state established during pairing was reused for all subsequent control actions across smart plugs, lighting devices, cameras, and hub mediated control.

*Authentication persistence over time*
Authentication persistence was evaluated by issuing control actions at increasing time intervals following initial pairing. For all evaluated devices, authentication artifacts remained valid for the entire duration of the experiments.
For smart plugs and the lighting device, authentication state remained unchanged for at least 72 hours, during which control actions were successfully issued without reauthentication. Application restarts did not invalidate authentication state in any observed case.
Camera devices exhibited similar persistence. Configuration and stream initiation requests remained authenticated after idle periods of up to 48 hours. No device required repairing or explicit reauthentication due solely to elapsed time.
Hub mediated control also showed long lived authentication behavior. Devices connected through the SmartThings hub accepted control commands without reauthentication after idle periods exceeding 72 hours. Across all devices, no explicit expiration or time based invalidation of authentication state was observed within the measurement window.

*Authentication behavior under network changes*
Authentication enforcement was further evaluated under routine network changes. Wireless disconnection and reconnection events were triggered ten times per device. In all cases, previously established authentication state remained valid after reconnection, and control actions succeeded without additional authentication exchanges. Address reassignment through DHCP renewal was performed five times per device. No device invalidated authentication state following address changes. Control requests issued after reassignment succeeded using the same authentication artifacts observed prior to the change. Router reboot events were conducted three times per device. After reconnection to the network, devices resumed normal operation using previously established authentication state. No repairing or reauthentication was required. Across all evaluated devices and network events, authentication persistence rate under network change conditions was 100 percent. Previously established authentication state remained valid after wireless reconnection, address reassignment, and router reboot across all evaluated devices.

*Replay feasibility and unauthorized control*
Replay feasibility was evaluated by reissuing previously captured control requests from a different host on the same local network. For the smart plugs and the lighting device, replay success rate was 100 percent. All replayed requests resulted in successful device actions without additional authentication. For camera devices, replay success depended on the action type. Configuration related requests exhibited a replay success rate of 80 percent. Stream initiation requests were accepted in 60 percent of replay attempts, with failures attributed to transport level session checks rather than authentication invalidation. Hub mediated control exhibited a replay success rate of 90 percent. Replayed requests resulted in successful control actions provided that message structure and authentication artifacts were preserved. Across all devices and action types, the overall replay success rate was 85 percent. Replay success was observed both under stable network conditions and after network changes, indicating that authentication artifacts were not bound to a specific controller or execution context.

### *Replay Attack Success Rate Across Device Categories*
To quantify replay feasibility across different device categories, replay experiments were aggregated and analyzed at the category level. For each device category, the replay success rate was computed as the proportion of replayed control requests that resulted in successful device actions without triggering re-authentication. This aggregation allows direct comparison of replay effectiveness across device types while abstracting away individual device instances. As shown in Fig 3, replay success rates range from 80% to 100% across the evaluated categories, with smart plugs and the smart lighting device exhibiting 100% success, IP cameras 80%, and the hub-mediated platform 90%. Fig 3 summarizes the replay success rates observed for smart plugs, the smart lighting device, IP cameras, and the hub-mediated platform.

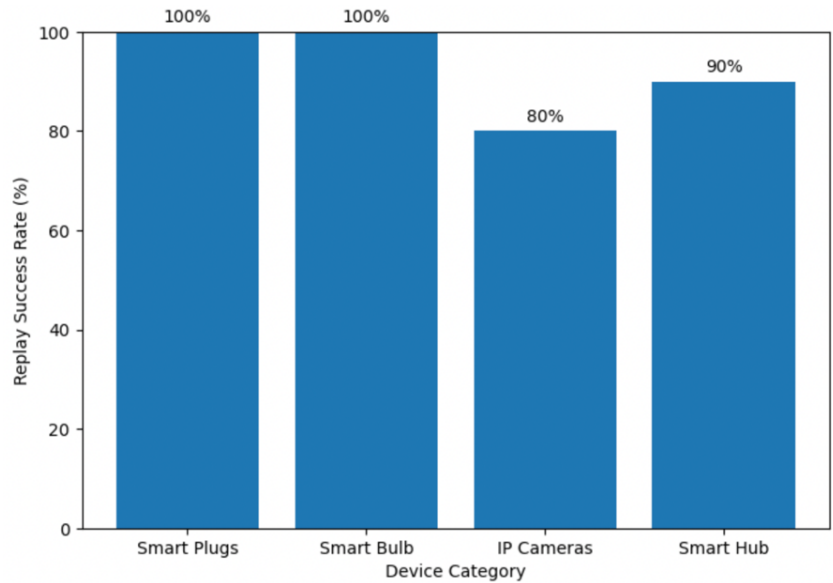

Fig 3. Replay success rate across smart home IoT device categories

### Discussion
The results reveal that authentication enforcement in the evaluated smart home IoT devices is dominated by long lived trust relationships established during initial pairing. Across all device categories, authentication artifacts were reused for the entirety of observed control actions, persisted over extended periods, and remained valid under routine network changes. While such behavior improves usability, it also introduces security risks that are not immediately visible to users. A key observation is the absence of time based or event driven invalidation of authentication state. For all evaluated devices, authentication artifacts remained valid for at least 48 to 72 hours, and in many cases longer, without explicit expiration. Neither application restarts nor extended idle periods triggered reauthentication. This suggests that authentication state is treated as a persistent capability rather than a session bound credential. From a security perspective, this design choice expands the window in which captured authentication artifacts can be reused by an attacker.

The lack of sensitivity to network context further amplifies this risk. Authentication state remained valid after wireless reconnection, address reassignment, and router reboot events with a success rate of 100 percent across all devices. This behavior indicates that authentication is not bound to network attributes such as IP address, connection instance, or link level identity. In residential environments where network access may be temporary or shared, this assumption effectively treats any host with network access as potentially trusted once authentication artifacts are obtained.

Replay experiments provide concrete evidence of the security impact of these design choices. For smart plugs and the lighting device, replay success rates reached 100 percent, indicating that previously observed authentication artifacts were sufficient to authorize control actions from a different host. Camera devices showed partial resistance for stream initiation, but configuration related requests were still accepted in a majority of replay attempts. Hub mediated control exhibited similarly high replay success rates. These findings demonstrate that authentication artifacts were not bound to a specific controller device or execution context.

Differences across device categories reflect varying priorities in system design rather than fundamentally different authentication models. Camera devices exhibited additional transport level checks for certain actions, which reduced replay success rates for stream initiation. However, the underlying authentication state remained persistent and reusable. This suggests that additional checks were implemented to protect high value operations rather than to strengthen authentication guarantees overall. Hub mediated control through the SmartThings ecosystem also relied on persistent authentication state, indicating that centralization alone does not address authentication binding or freshness.

These observations point to a broader pattern in smart home IoT design. Authentication mechanisms appear to be optimized for continuity of operation and minimal user friction. Reauthentication is avoided unless explicitly triggered by user action, such as device removal or factory reset. While this approach aligns with usability expectations, it implicitly assumes that local network access is benign and that authentication artifacts are not exposed. The results show that these assumptions do not hold under realistic threat models. Importantly, the observed weaknesses do not stem from broken cryptography or misconfigured transport security. All devices relied on encrypted communication channels and standard security protocols. The security risks arise instead from how authentication state is managed at the system level. This distinction matters, because it suggests that improving authentication security in smart home IoT systems requires changes in session management and trust modeling rather than simply stronger encryption.

The results also have implications for emerging interoperability ecosystems. Devices that participate in hub mediated control or standardized onboarding still exhibited persistent authentication behavior similar to standalone devices. This indicates that interoperability alone does not enforce stronger authentication semantics. Without explicit requirements for token rotation, context binding, or reauthentication triggers, standardized ecosystems may inherit the same weaknesses observed in proprietary systems.

Overall, the findings suggest that current smart home IoT authentication mechanisms provide continuity and convenience at the cost of reduced resistance to local network adversaries. Long lived authentication artifacts, lack of context binding, and high replay success rates collectively enable unauthorized control without requiring advanced attacker capabilities. Addressing these issues will require rethinking authentication as a continuously validated property rather than a one time pairing outcome.

**Conclusion**

This paper presented an empirical analysis of authentication enforcement in smart home IoT systems, focusing on how authentication state is established, maintained, and validated during normal operation. By examining real consumer devices under realistic residential conditions, the study evaluated authentication behavior as it is implemented in practice rather than as it is described in specifications or design documents. The results show that authentication mechanisms in the evaluated devices rely heavily on long lived trust relationships established during initial pairing. Authentication artifacts were consistently reused across control actions, persisted over extended periods without explicit expiration, and remained valid under routine network changes. Replay experiments further demonstrated that these artifacts could often be reused from a different host on the same local network to issue control commands successfully. These behaviors were observed across multiple device categories and ecosystems, indicating that they reflect common design patterns rather than isolated implementation choices.

Importantly, the identified security risks do not arise from broken cryptography or insecure transport protocols. All evaluated devices employed encrypted communication channels. Instead, the risks stem from system level authentication design decisions that favor continuity of operation and minimal user interaction. The absence of strong binding between authentication state and context, such as controller identity or network session, significantly increases exposure to local network adversaries with limited capabilities.

These findings highlight a gap between the security guarantees users may expect and the guarantees actually provided by deployed smart home IoT systems. While persistent authentication improves usability, it also expands the attack surface in environments where network access cannot be assumed to be fully trusted. Addressing this gap will require authentication mechanisms that incorporate stronger notions of freshness, context awareness, and revalidation, without imposing excessive burden on users.

Overall, this study underscores the need to treat authentication in smart home IoT systems as an ongoing security property rather than a one time pairing event. As smart homes continue to integrate deeper into everyday life, strengthening authentication enforcement will be essential to ensuring both usability and security in real world deployments.